\documentclass{article}%
\usepackage[dvips]{graphicx}
\usepackage{amsmath}
\usepackage{amsfonts}
\usepackage{amssymb}%
\begin{document}

\title{Tunnelling, Temperature and Taub-NUT Black Holes}
\author{Ryan Kerner\thanks{rkerner@sciborg.uwaterloo.ca} and R.B.
Mann\thanks{rbmann@sciborg.uwaterloo.ca}\\Department of Physics, University of Waterloo\\200 University Avenue West, Waterloo, Ontario N2L 3G1, Canada}
\maketitle

\begin{abstract}
We investigate quantum tunnelling methods for calculating black hole
temperature, specifically the null geodesic method of Parikh and Wilczek and
the Hamilton-Jacobi Ansatz method of Angheben et al. We consider application
of these methods to a broad class of spacetimes with event horizons, inlcuding
Rindler and non-static spacetimes such as Kerr-Newman and Taub-NUT. We obtain
a general form for the temperature of Taub-NUT-Ads black holes that is
commensurate with other methods. We examine the limitations of these methods
for extremal black holes, taking the extremal Reissner-Nordstrom spacetime as
a case in point.

\end{abstract}

\section{Introduction}

There are several methods for deriving Hawking radiation \cite{hawking}%
-\cite{Medved} and for calculating its temperature.\textit{\ \ }The original
method considered the creation of a black hole in the context of a collapse
geometry, calculating the Bogoliubov transformations between the initial and
final states of incoming and outgoing radiation. The more popular method of
analytic continuation to a Euclidean section (the Wick Rotation method)
emerged soon after\textbf{.} Relying on the methods of finite-temperature
quantum field theory, an analytic continuation $t\rightarrow i\tau$\ of the
black hole metric is performed and the periodicity of $\tau$\ (denoted by
$\beta$) is chosen in order to remove a conical singularity that would
otherwise be present at \ fixed points of the $U(1)$\ isometry generated by
$\partial/\partial\tau$\ (the event horizon in the original Lorentzian
section). \ The black hole is then considered to be in equilibrium with a
scalar field that has inverse temperature $\beta$\ at infinity.

Recently a semi-classical method of modeling Hawking radiation as a tunneling
effect was proposed \cite{krauswilczek1}-\cite{Medved}. \ This method
\ involves calculating the imaginary part of the action for the (classically
forbidden) process of \ s-wave emission across the horizon (first considered
by Kraus and Wilczek \cite{krauswilczek1}-\cite{krauswilczek3}), which in turn
is related to the Boltzmann factor for emission at the Hawking temperature.
Using the WKB approximation the tunneling probability for the classically
forbidden trajectory of the s-wave\textit{ }coming from inside to outside the
horizon is given by:%
\begin{equation}
\Gamma\propto\exp(-2\mathrm{Im}I)
\end{equation}
where $I$ is the classical action of the trajectory to leading order in
$\hslash$\ (here set equal to unity)\cite{krauswilczek}. \ Expanding the
action in terms of the particle energy, the Hawking temperature is recovered
at linear order. \ In other words for $2I=\beta E+O(E^{2})$ this gives%
\begin{equation}
\Gamma\thicksim\exp(-2I)\simeq\exp(-\beta E)
\end{equation}
which is the regular Boltzmann factor for a particle of energy $E$ where
$\beta$ is the inverse temperature of the horizon. \ The higher order terms
are a self-interaction effect resulting from energy conservation
\cite{krauswilczek},\cite{parikhwilczek}; however, for calculating the
temperature, expansion to linear order is all that is required. \ Two
different methods have been employed to calculate the imaginary part of the
action -- one used by Parikh and Wilczek \cite{parikhwilczek} \ and the other
by Angheben, Nadalini, Vanzo, and Zerbini \cite{radiation} (which is an
extension from the method used by Srinivasan and Padmanabhan \cite{Padman}).

The former method considers a null s-wave emitted from the black hole. Based
on previous work analyzing the full action in detail \cite{krauswilczek1}%
-\cite{krausvakkuri}, the only part of the action that contributes an
imaginary term is $\int_{r_{in}}^{r_{out}}p_{r}dr$, where $p_{r}$is the
momentum of the emitted null s-wave. \ Then by using Hamilton's equation and
knowledge of the null geodesics it is possible to calculate the imaginary part
of the action. \ We will refer to this approach as the null geodesic method.

The latter method involves consideration of an emitted\textbf{\ }scalar
particle, ignoring its self-gravitation, and assumes that its action satisfies
the relativistic Hamilton-Jacobi equation. From the symmetries of the metric
one picks an appropriate ansatz for the form of the action. \ We will refer to
this method as the Hamilton-Jacobi ansatz.\textbf{\ \ }

In this paper we examine these two methods in the context of a broader class
of spacetimes than has previously been studied. One of our prime motivations
is to understand the applicability of the method to stationary black hole
spacetimes such as the Kerr-Newman and Taub-NUT spacetimes.\textbf{\ }The
Taub-NUT metric is a generalization of the Schwarzschild metric and has played
an important role in the conceptual development of general relativity and in
the construction of brane solutions in string theory and M-theory.
\cite{NUTMtheory} The NUT charge plays the role of a magnetic mass, inducing a
topology in the Euclidean section at infinity that is a Hopf fibration of a
circle over a 2-sphere. \ \textquotedblleft A counter example to almost
anything\textquotedblright\ \cite{Misner}, \ \ Taub-NUT spaces have been of
particular interest in recent years because of the role they play in
furthering our understanding of the AdS-CFT correspondence
\cite{hawkinghuntpage},\cite{NutAds},\cite{Misnerentropy}.\textbf{\ }Along
these lines, the thermodynamics of various Taub-NUT solutions has been a
subject of intense study in recent years. Their entropy is not proportional to
the area of the event horizon and their free energy can sometimes be negative
\cite{hawkinghuntpage},\cite{Misnerentropy},\ \cite{TNresults},
\cite{hawkinghunt}. \ Solutions of Einstein equations with a negative
cosmological constant $\Lambda$\ and a nonvanishing NUT charge have a boundary
metric that has closed timelike curves. The behavior of quantum field theory
is significantly different in such spaces, and it is \ of interest to
understand how ADS/CFT works in these sorts of cases \cite{Taub}. \ 

All such thermodynamic calculations have thus far been carried out in the
Euclidean section, using Wick rotation methods. \ For most Taub-NUT spaces the
Lorentzian section has closed timelike curves. As a consequence, determination
of the temperature via the original method of Hawking -- while mathematically
clear -- is somewhat problematic in terms of its physical interpretation. It
is straightforward enough to analytically continue the time coordinate and
various metric parameters to render the metric Euclidean. \ Regularity
arguments then yield a periodicity for the time coordinate that can then be
interpreted as a temperature. \ However the Lorentzian analogue of this
procedure is less than clear, though it has been established that a
relationship between distinct analytic continuation methods exists
\cite{CrisRobb}. An independent method of computing the temperature associated
with event horizons in NUT-charged spacetimes is certainly desirable.

Our goal in this paper is to address this question, and to more generally
investigate the tunnelling approach outside of the spherically symmetric
ansatz. To this end, we compare the null geodesic method and the
Hamilton-Jacobi ansatz for obtaining the imaginary part of the action. We then
apply these methods to a variety of spacetimes, and derive a general formula
for the temperature from this method. \ We then consider specific cases of
interest, beginning with Rindler space and moving on to charged and rotating
black hole spacetimes. \ Turning to Taub-NUT spaces, we obtain a general
expression for the temperature for a subclass of Taub-NUT spacetimes without
closed timelike curves (CTCs) that we can compare to those obtained via Wick
rotation methods. We find agreement in all relevant cases.

Our paper is structured as follows. The next section will outline the two
methods, starting with a discussion and generalization of the null geodesic
method and followed by a discussion of the Hamilton-Jacobi ansatz. \ We
demonstrate that knowledge of the total mass or energy is not essential by
showing the direct application of these methods to Rindler spacetime. \ We
then apply these methods to stationary space-times, considering in turn the
Kerr-Newman class of metrics and then Taub-NUT spacetimes. In each case we
obtain results commensurate with other methods, concentrating in the latter
case on the subclass of Taub-NUT-AdS spacetimes that do not have closed
timelike curves \cite{Taub}.\ \ We finish with a preliminary discussion of
issues that occur when applying the method to extremal black holes,
concentrating on the specific case of the extremal Reissner-Nordstrom spacetime.

\section{Calculating the Imaginary Part of the Action For an Outgoing S-Wave}

\subsection{Null Geodesic Method}

We begin by reviewing the null geodesic method \cite{parikhwilczek}. \ The
general static spherically metric can be written in the form%

\begin{equation}
ds^{2}=-f(r)dt^{2}+\frac{dr^{2}}{g(r)}+r^{2}d\Omega^{2} \label{stat1}%
\end{equation}
which covers a broad range of black hole metrics. We want to write it in
Painlev\'{e} form \cite{Painleve} so that there is no singularity at the
horizon. \ This is easily accomplished via the transformation
\begin{equation}
t\rightarrow t-\int\sqrt{\frac{1-g\left(  r\right)  }{f\left(  r\right)
g\left(  r\right)  }}dr \label{paintrans}%
\end{equation}
\ yielding
\begin{equation}
ds^{2}=-f(r)dt^{2}+2\sqrt{f(r)}\sqrt{\frac{1}{g(r)}-1}drdt+dr^{2}+r^{2}%
d\Omega^{2} \label{painform}%
\end{equation}

This coordinate system has a number of interesting features. At any fixed time
the spatial geometry is flat. \ At any fixed radius the boundary geometry is
the same as that of the metric (\ref{stat1}).

The radial null geodesics for this metric correspond to
\begin{equation}
\dot{r}=\sqrt{\frac{f(r)}{g(r)}}\left(  \pm1-\sqrt{1-g(r)}\right)
\label{painrad}%
\end{equation}
where the plus/minus signs correspond to outgoing/ingoing null geodesics.

The basic idea behind this approach is to regard Hawking radiation as a
quantum tunnelling process. However unlike other tunnelling processes in which
two separated classical turning points are joined by a trajectory in imaginary
time, the tunnelling barrier is created by the outgoing particle itself, whose
trajectory is from the inside of the black hole to the outside, a classically
forbidden process. The probability of tunnelling is proportional to the
exponential of minus twice the imaginary part of the action for this process
in the WKB limit. \ Because of energy conservation, the radius of the black
hole shrinks as a function of the energy of the outgoing particle; in this
sense the particle creates its own tunnelling barrier.

In the spherically symmetric case the emitted particle is taken to be in an
outgoing s-wave mode, and so we use the plus sign in (\ref{painrad}). At the
horizon (where $g(r)=f(r)=0$) then $\dot{r}=0$ provided $\frac{f(r)}{g(r)}$ is
well defined there. The imaginary part of the action for an outgoing s-wave
from $r_{in}$ to $r_{out}$ is expressed as%
\begin{equation}
\operatorname{\mathrm{Im}}I=\operatorname{\mathrm{Im}}\int_{r_{in}}^{r_{out}%
}p_{r}dr=\operatorname{\mathrm{Im}}\int_{r_{in}}^{r_{out}}\int_{0}^{p_{r}%
}dp_{r}^{\prime}dr
\end{equation}
where $r_{in}$ and $r_{out}$ are the respective initial and final radii of the
black hole. The trajectory between these two radii is the barrier the particle
must tunnel through. \ 

We assume that the emitted s-wave has energy $\omega^{\prime}<<M$\ and that
the total energy of the space-time was originally $M$. Invoking conservation
of energy, to this approximation the s-wave moves in a background spacetime of
energy $M\rightarrow M-\omega^{\prime}$. \ In order to evaluate the integral,
we employ Hamilton's equation $\dot{r}=\frac{dH}{dp_{r}}|_{r}$\ to switch the
integration variable from momentum to energy ($dp_{r}=\frac{dH}{\dot{r}}$),
giving%
\begin{equation}
I=\int_{r_{in}}^{r_{out}}\int_{M}^{M-\omega}\frac{dr}{\dot{r}}dH=\int
_{0}^{\omega}\int_{r_{in}}^{r_{out}}\frac{dr}{\dot{r}}(-d\omega^{\prime})
\label{ssint}%
\end{equation}
where $dH=-d\omega^{\prime}$\ because total energy $H=M-\omega^{\prime}$\ with
$M$\ constant. \ Note that $\dot{r}$\ is implicitly a function of
$M-\omega^{\prime}$. For the special cases where this function is known (eg.
Schwarzschild) the integral in eq. (\ref{ssint}) can be solved exactly in
terms of $\omega$\ \cite{parikhwilczek}. \ Another generalization of the null
geodesic method \cite{extremal} spacetimes with a well defined ADM mass are
considered (since dependence of $M-\omega^{\prime}$\ is explicitly known) in
order to obtain self gravitation effects; for our considerations self
gravitation will be ignored\footnote{See ref. \cite{Medved} for a discussion
of self-gravitation effects in the context of the information-loss problem.}.

In general we can always perform a series expansion in $\omega$\ in order to
find the temperature. To first order this gives\textbf{\ \ }%
\begin{align}
I  &  =\int_{0}^{\omega}\int_{r_{in}}^{r_{out}}\frac{dr}{\dot{r}%
(r,M-\omega^{\prime})}(-d\omega^{\prime})=-\omega\int_{r_{in}}^{r_{out}}%
\frac{dr}{\dot{r}(r,M)}+O(\omega^{2})\nonumber\\
&  \simeq\omega\int_{r_{out}}^{r_{in}}\frac{dr}{\dot{r}(r,M)}
\label{nullaction}%
\end{align}
To proceed further we will need to estimate the last integral. First we note
that $r_{in}>r_{out}$\ because the black hole decreases in mass as the s-wave
is emitted; consequently the radius of the event horizon decreases. \ We
therefore write $r_{in}=r_{0}(M)-\epsilon$\ and $r_{out}=r_{0}(M-\omega
)+\epsilon$\ where $r_{0}(M)$\ denotes the location of the event horizon of
the original background space-time before the emission of particles.
\ Henceforth the notation $r_{0}$\ will be used to denote $r_{0}(M)$. Note
that with this generalization no explicit knowledge of the total energy or
mass is required since $r_{0}$\ is simply the radius of the event horizon
before any particles are emitted.

There is a pole at the horizon where $\dot{r}=0$. For a non-extremal black
hole $f^{\prime}(r_{0})$\ \ and $g^{\prime}(r_{0})$\ are both finite and
non-zero at the horizon, so for these cases $\frac{1}{\dot{r}}$\ only has a
simple pole at the horizon with a residue of $\frac{2}{\sqrt{f^{\prime}%
(r_{0})g^{\prime}(r_{0})}}$. \ Hence the imaginary part of the action will be%
\begin{equation}
\operatorname{\mathrm{Im}}I=\frac{2\pi\omega}{\sqrt{f^{\prime}(r_{0}%
)g^{\prime}(r_{0})}}+O(\omega^{2})
\end{equation}

Therefore the tunnelling probability is
\begin{equation}
\Gamma=\exp\left(  -2\mathrm{Im}I\right)  =\exp\left(  -\beta\omega\right)
\label{tunprob}%
\end{equation}
and so the Hawking temperature $T_{H}=\beta^{-1}$is%
\begin{equation}
T_{H}=\frac{\sqrt{f^{\prime}(r_{0})g^{\prime}(r_{0})}}{4\pi} \label{Thpar}%
\end{equation}

It is easy to confirm that for a Schwarzschild black hole the correct result
of $T_{H}=\frac{1}{8\pi M}$ follows. \ Situations in which the horizon does
not have a simple pole correspond to extremal black holes, and need to be
handled separately. One conceptual issue that arises when applying either the
Hamilton-Jacobi or null geodesic methods to the extremal case is due to the
fact that the model is dynamic, so emission of a neutral particle from the
black hole implies a naked singularity, in violation of cosmic censorship.
\ We will discuss the extremal case in Section 4.\textbf{\ \ \ }

\subsection{Hamilton-Jacobi Ansatz}

We next consider an alternate method for calculating the imaginary part of the
action making use of the Hamilton-Jacobi equation \cite{radiation}. \ We
assume that the action of the outgoing particle is given by the classical
action $I$ that satisfies the relativistic Hamilton-Jacobi equation%
\begin{equation}
g^{\mu\nu}\partial_{\mu}I\partial_{\nu}I+m^{2}=0 \label{HJac}%
\end{equation}
To leading order in the energy we can neglect the effects of the
self-gravitation of the particle.

For a metric of the form%
\begin{equation}
ds^{2}=-f(r)dt^{2}+\frac{dr^{2}}{g(r)}+C(r)h_{ij}dx^{i}dx^{j} \label{stat2}%
\end{equation}
the Hamilton-Jacobi equation (\ref{HJac}) is
\begin{equation}
-\frac{(\partial_{t}I)^{2}}{f(r)}+g(r)(\partial_{r}I)^{2}+\frac{h^{ij}}%
{C(r)}\partial_{i}I\partial_{j}I+m^{2}=0
\end{equation}
There exists a solution of the form
\begin{equation}
I=-Et+W(r)+J(x^{i})
\end{equation}
where
\[
\partial_{t}I=-E,\text{ \ \ \ \ \ }\partial_{r}I=W^{\prime}(r),\text{
\ \ \ \ }\partial_{i}I=J_{i}
\]
and that the $J_{i}$'s are constant. \ Solving for $W(r)$ yields%
\begin{equation}
W(r)=\int\frac{dr}{\sqrt{f(r)g(r)}}\sqrt{E^{2}-f(r)(m^{2}+\frac{h^{ij}%
J_{i}J_{j}}{C(r)})} \label{wdef}%
\end{equation}
(for an outgoing particle) and the imaginary part of the action can only come
from the pole at the horizon. \ It is important to parameterize in terms of
the proper spatial distance in order to get the correct result
\cite{radiation}. \ 

For the first method the Painlev\'{e} coordinate $r$ was the proper spatial
distance. In this case the proper spatial distance between any two points at
some fixed $t$ is given by%
\begin{equation}
d\sigma^{2}=\frac{dr^{2}}{g(r)}+C(r)h_{ij}dx^{i}dx^{j}%
\end{equation}
As with the null geodesic method we are only concerned with radial rays, and
so the only proper spatial distance we are concerned with is radial%

\[
d\sigma^{2}=\frac{dr^{2}}{g(r)}
\]

Employing the near horizon approximation
\begin{align}
f(r)  &  =f^{\prime}(r_{0})(r-r_{0})+...\\
g(r)  &  =g^{\prime}(r_{0})(r-r_{0})+...\nonumber
\end{align}
we find that%

\begin{equation}
\sigma=\int\frac{dr}{\sqrt{g(r)}}\simeq2\frac{\sqrt{r-r_{0}}}{\sqrt{g^{\prime
}(r_{0})}}%
\end{equation}
is the proper radial distance. \ So for particles emitted radially%
\begin{align}
W(\sigma)  &  =\frac{2}{\sqrt{g^{\prime}(r_{0})f^{\prime}(r_{0})}}\int
\frac{d\sigma}{\sigma}\sqrt{E^{2}-\frac{\sigma^{2}}{4}g^{\prime}%
(r_{0})f^{\prime}(r_{0})\left(  m^{2}+\frac{h^{ij}J_{i}J_{j}}{C(r_{0}%
)}\right)  }\nonumber\\
&  =\frac{2\pi iE}{\sqrt{g^{\prime}(r_{0})f^{\prime}(r_{0})}} \label{FormW}%
\end{align}
and from this point the computation is the same as for the previous method,
yielding
\begin{equation}
T_{H}=\frac{\sqrt{f^{\prime}(r_{0})g^{\prime}(r_{0})}}{4\pi} \label{Thvan}%
\end{equation}
for the temperature. \ 

\section{Applications}

\subsection{Rindler Space}

We first illustrate how these methods apply for the horizon of an accelerated
observer. We shall employ different coordinate systems for 2D Rindler space to
show that the same temperature results from applying the two tunneling methods directly.

The forms of the Rindler metric being used are:%

\begin{align}
ds^{2}  &  =-(a^{2}x^{2}-1)dt^{2}+\frac{a^{2}x^{2}}{a^{2}x^{2}-1}%
dx^{2}\label{rindler1}\\
ds^{2}  &  =-a^{2}x^{2}dt^{2}+dx^{2} \label{rindler2a}%
\end{align}
where $a$ is the proper acceleration of the hyperbolic observer. Here there is
no well defined total mass or energy as with Schwarzschild, but there are well
defined horizons. \ The metric (\ref{rindler1}) locates the horizon at
$x=\frac{1}{a}$, whereas for the metric (\ref{rindler2a}) it is at $x=0$\ .\ 

We consider a null particle to be emitted from the Rindler horizon, and it is
reasonable to assume the emitted particle will have a Hamiltonian associated
with it. However providing an explicit definition for the total energy of the
space-time is less than clear, though it has been claimed recently
\cite{rindler2} that one can associate a surface energy density $\sigma
=\frac{a}{4\pi}$\ with a Rindler horizon and a total energy $E=\frac{1}{4a}%
$\ with the spacetime. \ \ In the context of the null geodesic method \ we
expect that the Hamiltonian of the space-time will correspond to the total
energy $E$\ (perhaps with respect to some reference energy via a limiting
procedure)\textbf{\ }so as long as the emitted particles have $\omega
<<\frac{1}{4a}$, in which case the method is applicable. \ \ We shall proceed
under the assumption that we can use Hamilton's equation and follow through
the derivation for the null geodesic method as before.\ We shall find that
these assumptions are justified a-posteriori.\ \textbf{\ }

The null geodesics for (\ref{rindler2a}) in the $x$-direction are given by%

\[
\dot{x}=\pm ax
\]
and so
\[
\operatorname{\mathrm{Im}}I=\omega\int_{x_{in}}^{x_{out}}\frac{dx}{ax}%
=\frac{\pi\omega}{a}
\]
yielding a temperature of%
\[
T_{H}=\frac{a}{2\pi}=\frac{a\kappa}{2\pi}
\]
where the surface gravity at the horizon is $\kappa=1$ \cite{rindler}. \ 

We now employ the Hamilton-Jacobi ansatz\footnote{For earlier work in the
Rindler context along these lines see ref.\cite{Padman}.} for the Rindler
metric (\ref{rindler1}).

Here $f=a^{2}x^{2}-1,g=\frac{a^{2}x^{2}-1}{a^{2}x^{2}}$ and at the horizon
$f^{\prime}(1)=g^{\prime}(1)=2a$ so using (\ref{FormW})
\[
W=\frac{E\pi i}{a}
\]
again giving a temperature of $T_{H}=\frac{a}{2\pi}$.

We see that we can recover the expected value for the temperature of Rindler
space given our assumptions. \ This could perhaps be regarded further evidence
that a total energy $E=\frac{1}{4a}$\ can be associated with Rindler space.

\subsection{Charged-Kerr Black Hole}

We consider next the Kerr-Newman solution. \ The Kerr-Newman metric and vector
potential are given by\textit{\ }%
\begin{align}
ds^{2}  &  =-f(r,\theta)dt^{2}+\frac{dr^{2}}{g(r,\theta)}-2H(r,\theta
)dtd\phi+K(r,\theta)d\phi^{2}+\Sigma(r,\theta)d\theta^{2}\nonumber\\
A_{a}  &  =-\frac{er}{\Sigma(r)}[(dt)_{a}-a^{2}\sin^{2}\theta(d\phi)_{a}]\\
f(r,\theta)  &  =\frac{\Delta(r)-a^{2}\sin^{2}\theta}{\Sigma(r,\theta
)},\nonumber\\
g(r,\theta)  &  =\frac{\Delta(r)}{\Sigma(r,\theta)},\nonumber\\
H(r,\theta)  &  =\frac{a\sin^{2}\theta(r^{2}+a^{2}-\Delta(r))}{\Sigma
(r,\theta)}\nonumber\\
K(r,\theta)  &  =\frac{(r^{2}+a^{2})^{2}-\Delta(r)a^{2}\sin^{2}\theta}%
{\Sigma(r,\theta)}\sin^{2}(\theta)\nonumber\\
\Sigma(r,\theta)  &  =r^{2}+a^{2}\cos^{2}\theta\nonumber\\
\Delta(r)  &  =r^{2}+a^{2}+e^{2}-2Mr\nonumber
\end{align}
We assume a non-extremal black hole so that $M^{2}>a^{2}+e^{2}$\ so that there
are two horizons at $r_{\pm}=M\pm\sqrt{M^{2}-a^{2}-e^{2}}$.

There is a technical issue in applying these methods because the metric
functions depend on the angle $\theta$. \ In order to account for this we can
no longer just look a generic spherical wave; instead we will examine rings of
emitted photons for arbitrary fixed $\theta=\theta_{0}$. \ In the end we will
discover our temperature is independent of $\theta_{0}$\ (as it should be). \ 

A naive first attempt utilizing the null geodesic method would be to consider
the transformation
\[
dt=dT-\sqrt{\frac{1-g(r,\theta_{0})}{g(r,\theta_{0})f(r,\theta_{0})}}dr
\]
This gives the equation
\begin{align}
ds^{2}  &  =-f(r,\theta_{0})dT^{2}+2\sqrt{f(r,\theta_{0})}\sqrt{\frac
{1}{g(r,\theta_{0})}-1}drdT+dr^{2}\nonumber\\
&  -2Hd\phi(dT-\frac{\sqrt{\frac{1}{g(r,\theta_{0})}-1}}{\sqrt{f(r,\theta
_{0})}}dr)+Kd\phi^{2}%
\end{align}
whose radial null geodesics correspond to\textit{\ }%
\begin{equation}
\dot{r}=\sqrt{\frac{f(r,\theta_{0})}{g(r,\theta_{0})}}\left(  \pm
1-\sqrt{1-g(r,\theta_{0})}\right)
\end{equation}

There remain divergences in the $dtdr$\ and $drd\phi$\ terms at the horizon,
and $\frac{f(r,\theta_{0})}{g(r,\theta_{0})}$\ is not well behaved there. Only
for $\sin\theta_{0}=0$\ are these eliminated. \ Restricting further the
calculation to $\theta_{0}=0$\ or $\pi$\ (in which case $\frac{f}{g}=1$), the
outgoing radial null geodesics along the $z$\ axis are
\begin{equation}
{\dot{r}=1-}\sqrt{1-g(r,\theta_{0})|_{\sin\theta_{0}=0}}%
\end{equation}
which yields
\[
I=\omega\int_{r_{out}}^{r_{in}}\frac{dr}{\dot{r}}=\frac{2\pi\omega}{g^{\prime
}(r_{+},\theta_{0})|_{\sin\theta=0}}=2\pi\omega\frac{r_{+}^{2}+a^{2}}%
{2(r_{+}-M)}
\]
for the imaginary part of the action{}. This in turn results in the
temperature%
\begin{equation}
T_{H}=\frac{1}{2\pi}\frac{r_{+}-M}{r_{+}^{2}+a^{2}}=\frac{1}{2\pi}\frac
{(M^{2}-a^{2}-e^{2})^{\frac{1}{2}}}{2M(M+(M^{2}-a^{2}-e^{2})^{\frac{1}{2}%
})-e^{2}}%
\end{equation}
which is the same as the found for the Kerr-Newman black hole by other means.

The restriction to two specific values of $\theta_{0}$\ is because of the
presence of the ergosphere. \ The calculation breaks down because
$f(r,\theta)$\ is actually negative elsewhere at the horizon (i.e. inside the
ergosphere) and $\partial_{T}$\ is not properly timelike there . \ The two
values $\theta_{0}=0$\ or $\pi$\ correspond to where the event horizon and
ergosphere coincide. \ 

To address this issue, we note that the original charged Kerr metric can be
rewritten as
\begin{align}
ds^{2}  &  =-F(r,\theta)dt^{2}+\frac{dr^{2}}{g(r,\theta)}+K(r,\theta
)(d\phi-\frac{H(r,\theta)}{K(r,\theta)}dt)^{2}+\Sigma(r)d\theta^{2}%
\label{kerr}\\
F(r,\theta)  &  =f(r,\theta)+\frac{H^{2}(r,\theta)}{K(r,\theta)}=\frac
{\Delta(r)\Sigma(r,\theta)}{(r^{2}+a^{2})^{2}-\Delta(r)a^{2}\sin^{2}\theta
}\nonumber
\end{align}
where at the horizon%

\[
\frac{H(r_{+},\theta)}{K(r_{+},\theta)}\mathbf{=}\frac{a}{r_{+}^{2}+a^{2}%
}\mathbf{=\Omega}_{H}
\]
So the metric near the horizon for fixed $\theta=\theta_{0}$\ is%
\begin{equation}
ds^{2}=-F_{r}(r_{+},\theta_{0})(r-r_{+})dt^{2}+\frac{dr^{2}}{g_{r}%
(r_{+},\theta_{0})(r-r_{+})}+K(r_{+},\theta_{0})(d\phi-\frac{H(r_{+}%
,\theta_{0})}{K(r_{+},\theta_{0})}dt)^{2} \label{nearkerr}%
\end{equation}
and defining $d\chi=d\phi-\frac{H(r_{+},\theta_{0})}{I(r_{+},\theta_{0})}dt$.%
\begin{equation}
ds^{2}=-F_{r}(r_{+},\theta_{0})(r-r_{+})dt^{2}+\frac{dr^{2}}{g_{r}%
(r_{+},\theta_{0})(r-r_{+})}+K(r_{+},\theta_{0})(d\chi)^{2} \label{chikerr}%
\end{equation}

The metric (\ref{chikerr}) is well-behaved for all $\theta_{0}$\ and is of the
same form as (\ref{stat1})\ with $f(r)=F_{r}(r_{+},\theta_{0})(r-r_{+})$\ and
$g(r)=g_{r}(r_{+},\theta_{0})(r-r_{+})$\ . Hence we easily obtain the final
result (\ref{Thpar})%
\[
T_{H}=\frac{\sqrt{F_{r}(r_{+},\theta_{0})g_{r}(r_{+},\theta_{0})}}{4\pi}
\]

Explicit calculation of $F_{r}(r_{+},\theta_{0})$\ and $g_{r}(r_{+},\theta
_{0})$\ yields%
\begin{align*}
g_{r}(r_{+},\theta_{0})  &  =\frac{\Delta_{r}(r_{+})}{\Sigma(r_{+},\theta
_{0})}=\frac{2r_{+}-2M}{r_{+}^{2}+a^{2}\cos^{2}(\theta_{0})}\\
F_{r}(r_{+},\theta_{0})  &  =\frac{\Delta_{r}(r_{+})\Sigma(r_{+},\theta_{0}%
)}{(r_{+}^{2}+a^{2})^{2}}=\frac{(2r_{+}-2M)(r_{+}^{2}+a^{2}\cos^{2}(\theta
_{0}))}{(r_{+}^{2}+a^{2})^{2}}%
\end{align*}
Although $F_{r}(r_{+},\theta_{0})$\ and $g_{r}(r_{+},\theta_{0})$\ each depend
on $\theta_{0}$, their product
\[
F_{r}(r_{+},\theta_{0})g_{r}(r_{+},\theta_{0})=\frac{(2r_{+}-2M)^{2}}%
{(r_{+}^{2}+a^{2})^{2}}%
\]
\ is independent of this quantity. \ Hence the temperature is
\[
T_{H}=\frac{1}{2\pi}\frac{r_{+}-M}{r_{+}^{2}+a^{2}}=\frac{1}{2\pi}\frac
{(M^{2}-a^{2}-e^{2})^{\frac{1}{2}}}{2M(M+(M^{2}-a^{2}-e^{2})^{\frac{1}{2}%
})-e^{2}}%
\]
for any angle.

We turn next to the Hamilton-Jacobi method to find the temperature. The action
is assumed to be of the form%
\[
I=-Et+J\phi+W(r,\theta_{0})
\]
and rewriting this in terms of $\chi(r_{+})=$\ $\phi-\Omega_{H}t$\ we find%
\[
I=-(E-\Omega_{H}J)t+J\chi+W(r,\theta_{0})
\]
where it is assumed that $E-\Omega_{H}J>0$. This demonstrates a nuance
overlooked in the null geodesic method; \ the transformation to $\chi
$\ implies that $E$\ should be replaced by $E-\Omega_{H}J$\ for the emitted
particle. \ The reason for this is the presence of the ergosphere. \ The
Killing field that is timelike everywhere is $\chi=\partial_{t}+\Omega
_{H}\partial_{\phi}$. A particle can escape to infinity only if $p_{a}\chi
^{a}<0$\ , and so $-E+\Omega_{H}J<0$\ where $E$\ and $J$\ are the energy and
angular momentum of the particle.

Employing the metric in the near horizon form (\ref{nearkerr}), the final
result for $W(r,\theta_{0})$\ is the same as (\ref{FormW}) with $E$\ replaced
by $E-$\ $\Omega_{H}J$:%
\begin{equation}
W(r,\theta_{0})=\frac{2\pi i(E-\Omega_{H}J)}{\sqrt{F_{r}(r_{+},\theta
_{0})g_{r}(r_{+},\theta_{0})}}=(E-\Omega_{H}J)\frac{\pi i(r_{+}^{2}+a^{2}%
)}{(r_{+}-M)}%
\end{equation}
again yielding the temperature over the full surface of the Black Hole%
\[
T_{H}=\frac{1}{2\pi}\frac{r_{+}-M}{r_{+}^{2}+a^{2}}=\frac{1}{2\pi}\frac
{(M^{2}-a^{2}-e^{2})^{\frac{1}{2}}}{2M(M+(M^{2}-a^{2}-e^{2})^{\frac{1}{2}%
})-e^{2}}
\]
in full agreement with the previous method and with Euclidean space
techniques.\textbf{\ \ \ }

\subsection{Taub-NUT-AdS}

The general Taub-NUT-AdS solutions with cosmological constant $\Lambda
=-3/\ell^{2}$ are given by \cite{Taub}%

\begin{equation}
ds^{2}=-F(r)(dt+4n^{2}f_{k}^{2}(\frac{\theta}{2})d\varphi)^{2}+\frac{dr^{2}%
}{F(r)}+(r^{2}+n^{2})(d\theta^{2}+f_{k}^{2}(\theta)d\varphi^{2}) \label{taub1}%
\end{equation}
where
\begin{equation}
F(r)=k\frac{r^{2}-n^{2}}{r^{2}+n^{2}}+\frac{-2Mr+\frac{1}{\ell^{2}}%
(r^{4}+6n^{2}r^{2}-3n^{4})}{r^{2}+n^{2}}%
\end{equation}
and $k$ is a discrete parameter that takes the values $1,0,-1$ and defines the
form of the function $f_{k}(\theta)$
\begin{equation}
f_{k}(\theta)=\left\{
\begin{array}
[c]{c}%
\sin\theta\qquad\text{for k}=1\\
\theta\qquad\text{for k}=0\\
\sinh\theta\qquad\text{for k}=-1
\end{array}
\right.
\end{equation}

One of the interesting properties of Taub-NUT spaces is the existence of
closed timelike curves (CTCs) \cite{Misner}. For these cases it is not clear
how to apply the null geodesic method, since the emission of an s-wave
particle would have to recur in a manner consistent with the presence of CTCs.

However\ there exists a special subclass of Hyperbolic Taub-NUT solutions (for
$4n^{2}/\ell^{2}\leq1$\ ) that do not contain CTCs. \ A discussion of Taub-NUT
space and the special cases without CTCs appears in the appendix. We shall
consider these cases in what follows.

The temperature can be successfully calculated using the metric in the
following form:%
\begin{equation}
ds^{2}=-Hdt^{2}+\frac{dr^{2}}{F}+G(d\varphi-\frac{F4nf_{k}^{2}(\frac{\theta
}{2})}{G}dt)^{2}+(r^{2}+n^{2})d\theta^{2} \label{taub2}%
\end{equation}
where:
\begin{align}
H(r,\theta)  &  =(F+F^{2}\frac{16n^{2}f_{k}^{4}(\frac{\theta}{2})}{G})\\
G(r,\theta)  &  =4f_{k}^{2}(\frac{\theta}{2})\left(  r^{2}+n^{2}-f_{k}%
^{2}(\frac{\theta}{2})(4n^{2}F+k(r^{2}+n^{2}))\right)
\end{align}
As before, we will consider rings at constant $\theta_{0}$\ and use the near
horizon approximation.

\bigskip At the horizon%
\[
\frac{G(r_{+},\theta_{0})}{f_{k}^{2}(\frac{\theta_{0}}{2})}=\left\{
\begin{array}
[c]{c}%
4\left(  (r_{+}^{2}+n^{2})\cosh^{2}(\frac{\theta_{0}}{2})\right)  ,\text{
\ \ }k=-1\\
4(r_{+}^{2}+n^{2}),\text{ \ \ \ \ \ \ \ \ \ \ \ \ \ \ \ \ \ \ \ \ \ \ \ \ \ }%
k=0\\
4\left(  (r_{+}^{2}+n^{2})\cos^{2}(\frac{\theta_{0}}{2})\right)  ,\text{
\ \ \ }k=1
\end{array}
\right.
\]
Only when $k=1$\ (for which CTCs are present) and $\theta_{0}=\pi$\ (i.e. when
$\cos(\frac{\theta_{0}}{2})=0$) \ are there any potential divergences at the
horizon. \ Since%
\[
H_{r}(r_{+},\theta_{0})=F_{r}(r_{+})
\]
the metric near the horizon for fixed $\theta=\theta_{0}$\ is%

\begin{align}
ds^{2}  &  =-F_{r}(r_{+})(r-r_{+})dt^{2}+\frac{dr^{2}}{F_{r}(r_{+})(r-r_{+}%
)}\nonumber\\
&  +G(r_{+},\theta_{0})(d\varphi-\frac{F_{r}(r_{+})4nf_{k}^{2}(\frac{\theta
}{2})}{G(r_{+},\theta_{0})}(r-r_{+})dt)^{2}\\
&  =-F_{r}(r_{+})(r-r_{+})dt^{2}+\frac{dr^{2}}{F_{r}(r_{+})(r-r_{+})}%
+G(r_{+},\theta)d\varphi^{2}%
\end{align}
Notice that defining $\chi=\varphi-\Omega_{H}t$\ as with the charged Kerr case
is pointless since $\Omega_{H}=0$. \ \ From this point the steps are the same
as for the general procedures outlined for either the null-geodesic method or
the Hamilton-Jacobi ansatz. Inserting\textbf{\ }this into the final result for
temperature (either (\ref{Thpar}) or (\ref{Thvan})) yields\bigskip%

\begin{equation}
T_{H}=\frac{F_{r}(r_{+})}{4\pi} \label{Temptaub}%
\end{equation}
which is the same form found using the Wick rotation method
\cite{CrisRobb,Taub}.

To demonstrate this is straightforward. Consider the hyperbolic case
($k=-1$)\textbf{.} The mass parameter can be written in terms of the other
metric parameters upon recognition that $F\left(  r_{+}\right)  =0$\ yielding%
\[
M=\frac{r_{+}^{4}+(6n^{2}-\ell^{2})r_{+}^{2}-n^{2}(3n^{2}-\ell^{2})}{2\ell
^{2}r_{+}}
\]

Using this mass in (\ref{Temptaub}) yields an expression for the hyperbolic
Taub-NUT temperature of%
\begin{equation}
T_{H}=\frac{4\pi\ell^{2}r_{+}}{3(r_{+}^{2}+n^{2})-\ell^{2}}%
\end{equation}

Comparing this to the result (\cite{Taub}) for the hyperbolic Taub-NUT
temperature obtained from Wick rotation methods
\begin{equation}
T_{H}=\frac{4\pi\ell^{2}r_{+}}{3(r_{+}^{2}-N^{2})-\ell^{2}}=\frac{F_{r}%
(r_{+})}{4\pi}%
\end{equation}
(where $N$\ is the Wick rotated NUT charge) we obtain agreement upon
recognizing that $n^{2}=-N^{2}$\ due to analytic continuation. \ Note however
that there is an implicit analytic continuation in the definition of
\textbf{\ }$r_{+}$, since $F\left(  r_{+},n\right)  \rightarrow F\left(
r_{+},iN\right)  $ \cite{CrisRobb}.

We close by commenting that although we considered only the $k=-1$\ case to
avoid problems with CTCs, both the $k=0,1$\ cases can be formally carried
through, yielding the result (\ref{Temptaub}). In the context of the null
geodesic method this situation could perhaps be interpreted by noting that
Hawking radiation yields a thermal bath of particles, whose existence can
statistically be reconciled with the presence of CTCs. \ In the context of the
Hamilton-Jacobi ansatz the physical interpretation is less problematic
provided the classical action for the particle can be considered to obey the
Hamilton-Jacobi equation in the presence of CTCs. \ Our results suggest
a-posteriori the answer is yes, but the matter merits further study.
\textbf{\ }In this context we note recent work \cite{Holzegel}\ demonstrating
that there are no SU(2)-invariant (time-dependent) tensorial perturbations of
asymptotically flat Lorentzian Taub-NUT space, calling into question the
possibility that a physically sensible thermodynamics can be associated to
Lorentzian Taub-NUT spaces without cosmological constant. \ Whether or not
such results extend to Taub-NUT spaces without CTCs is an interesting question.

\section{Extremal Black Holes}

Extremal black holes need to be treated separately from the other
generalizations, since\ the integrand no longer has a single pole. The general
results derived above are no longer valid and even the self gravitating terms
may play a very important role. \ One of the properties that occurs in
extremal case is the presence of a divergent real component in the action.
\ Although such a term does not contribute to the imaginary part of the
action, this may be an indication that the tunnelling approach is breaking
down and the calculation is becoming pathological. \ Unlike the Wick-rotation
method, which involves finding an equilibrium temperature, the tunnelling
approach describes a dynamical system. \ In this latter context when a black
hole is extremal the possibility exists that an emitted neutral particle may
cause the creation of a naked singularity, in violation of cosmic censorship. \ 

Such a pathological situation would be prevented if the tunnelling barrier had
infinite height. However we do not find this to be the case, and an evaluation
of the imaginary part of the action yields a finite temperature. This is
consistent with the proposal that extremal black holes can be in thermal
equilibrium at any temperature \cite{HHR}.

For concreteness, we shall consider the particular case of the
Reissner-Nordstrom metric , though we note that a diverging real component has
also been seen to occur with the extremal GHS solution \cite{radiation}.

\subsection{Extremal Reissner-Nordstrom}

The Reissner-Nordstrom space-time is described by the metric%
\begin{equation}
ds^{2}=-(1-\frac{2M}{r}+\frac{Q^{2}}{r^{2}})dt^{2}+\frac{dr^{2}}{(1-\frac
{2M}{r}+\frac{Q^{2}}{r^{2}})}+r^{2}d\Omega^{2}%
\end{equation}
The black hole is non-extremal when $M^{2}>Q^{2}$\ and extremal when $Q=M$.
\ For the non-extremal case when the tunnelling approach yields a temperature of%

\begin{equation}
T_{H}=\frac{1}{2\pi}\frac{\sqrt{M^{2}-Q^{2}}}{(M+\sqrt{(M^{2}-Q^{2})})^{2}}
\label{thrn}%
\end{equation}
using either of (\ref{Thpar}) or (\ref{Thvan}). Note that the limit
$Q\rightarrow M$\ gives a temperature of zero.

For the Reissner-Nordstrom case self gravitating effects have been calculated
exactly \cite{parikhwilczek} and the full emission rate is%

\begin{equation}
\Gamma\sim e^{-2I}=e^{-2\pi\left(  2\omega(M-\frac{\omega}{2})-(M-\omega
)\sqrt{(M-\omega)^{2}-Q^{2}}+M\sqrt{M^{2}-Q^{2}}\right)  } \label{RNemission}%
\end{equation}
Expanding this emission rate in powers of $\omega$ yields the temperature
(\ref{thrn}) to leading order. Note that setting $Q=M$ yields a contradictory
result, since the second term in the exponent becomes imaginary. This
unphysical situation corresponds to an extremal black hole emitting a
particle, a situation in violation of cosmic censorship.

Consider a nearly extremal black hole that emits a particle so that the
resulting black hole is extremal. This corresponds to substitution of
$Q=(M-\omega)$ where the black hole emits a null particle of energy $\omega$.
Insertion of this value of $Q$ into (\ref{thrn}) yields
\begin{equation}
T_{H}=\frac{1}{2\pi}\frac{\sqrt{M^{2}-(M-\omega)^{2}}}{(M+\sqrt{(M^{2}%
-(M-\omega)^{2})})^{2}}=\frac{1}{2\pi}\frac{\sqrt{2M\omega}}{M^{2}}+O(\omega)
\label{Tex1}%
\end{equation}
Comparing this to the temperature obtained from the emission rate using
(\ref{RNemission}) gives
\[
\Gamma=e^{-2\pi\left(  2\omega(M-\frac{\omega}{2})+M\sqrt{M^{2}-(M-\omega
)^{2}}\right)  }=e^{-2\pi\left(  M\sqrt{2M\omega}+2\omega M+O(\omega^{\frac
{3}{2}})\right)  }
\]
From the definition (\ref{tunprob}) we find that the temperature that is
$O(\sqrt{\omega})$ and again approaches zero the closer the original black
hole is to extremality. Explicitly
\begin{equation}
T=\frac{1}{2\pi}\frac{\omega}{M\sqrt{2M\omega}}=\frac{1}{4\pi}\frac
{\sqrt{2M\omega}}{M^{2}} \label{Tex2}%
\end{equation}
which differs from the value given in (\ref{Tex1}) by a factor of $1/2$.
\ This discrepancy arises due to an inappropriate expansion implicitly used in
obtaining (\ref{Tex1}), which assumes that $\omega<<\frac{M^{2}-Q^{2}}{2M}$\ ,
an invalid assumption for $Q=(M-\omega)$\textbf{.} In this context we note
earlier work demonstrating that the transition probability of emitting such a
particle that will make the black hole extremal is zero \cite{krauswilczek3}.

We obtain a temperature that depends on the energy of the emitted particle. We
pursue the extremal case further by considering a direct attempt to find the
temperature from the metric in its extremal form%
\begin{equation}
ds^{2}=-(1-\frac{M}{r})^{2}dt^{2}+\frac{dr^{2}}{(1-\frac{M}{r})^{2}}%
+r^{2}d\Omega^{2}%
\end{equation}
Using the Hamilton-Jacobi Ansatz as a first attempt at the calculation yields
only a diverging real component. i.e.\textbf{\ }%

\[
f(r)=g(r)=\frac{1}{M^{2}}(r-M)^{2}+O((r-M)^{3})
\]
\textbf{\ }so that%
\begin{align*}
\sigma &  =\int\frac{dr}{\sqrt{g(r)}}=M\int\frac{dr}{(r-M)}\simeq M\ln(r-M)\\
r-M  &  =e^{\frac{\sigma}{M}}%
\end{align*}
where $M<r<\infty$\ implies that bounds on $\sigma$\ are now $-\infty
<\sigma<\infty$.\ \ Rather than considering an observer at infinity, we will
consider an observer outside the horizon at some $r_{1}$ corresponding to
$\sigma(r_{1})$. \ From (\ref{wdef})%
\begin{align}
W(r)  &  =\int\frac{dr}{\frac{1}{M^{2}}(r-M)^{2}}\sqrt{E^{2}-\frac{1}{M^{2}%
}(r-M)^{2}(m^{2}+\frac{h^{ij}J_{i}J_{j}}{C(r)})}\nonumber\\
&  =M\int_{-\infty}^{\sigma(r_{1})}\frac{d\sigma}{e^{\frac{\sigma}{M}}}%
\sqrt{E^{2}-\frac{1}{M^{2}}e^{2\frac{\sigma}{M}}(m^{2}+\frac{h^{ij}J_{i}J_{j}%
}{C(r_{0})})}\nonumber\\
&  =M\int_{-\infty}^{\sigma(r_{1})}d\sigma\sqrt{E^{2}e^{-2\frac{\sigma}{M}%
}-\frac{1}{M^{2}}(m^{2}+\frac{h^{ij}J_{i}J_{j}}{C(r_{0})})}%
\end{align}
For convenience we choose $\sigma(r_{1})$\ so that the term under the root is
never negative. \ This integral is diverging and real, suggesting that no
particles are emitted \cite{radiation}. \ \ However this result is suspect in
that it may be contingent on employing the near horizon approximation in the
early stages of this method.

\bigskip

We turn next to the null geodesic method. The outward radial null geodesic is
given by\textit{\ }%
\begin{align}
\dot{r}  &  =1-\sqrt{1-(1-\frac{M}{r})^{2}}\\
&  =\frac{1}{2M^{2}}(r-M)^{2}-\frac{1}{M^{3}}(r-M)^{3}+O((r-M)^{4})
\end{align}
\textbf{\ }Insertion of this into (\ref{nullaction}) yields%
\begin{align}
\operatorname{Im}I  &  \simeq\operatorname{Im}\left[  -\omega\int_{0}^{\pi
}\frac{\epsilon ie^{i\theta}d\theta}{\frac{1}{2M^{2}}\epsilon^{2}e^{2i\theta
}(1+\frac{2}{M}\epsilon e^{i\theta})}\right] \nonumber\\
&  =-2\omega M^{2}\operatorname{Im}\left[  \int_{0}^{\pi}(\frac{i}{\epsilon
e^{i\theta}}+\frac{2i}{M(1+\frac{2}{M}\epsilon e^{i\theta})})d\theta\right]
\nonumber\\
&  =\operatorname{Im}\left[  O(\frac{1}{\epsilon})+4M\omega\lbrack
\ln(e^{-i\theta}+\frac{2\epsilon}{M})]|_{0}^{\pi}\right] \nonumber\\
&  =4M\omega\operatorname{Im}\left[  \ln\left(  \frac{-1+\frac{2\epsilon}{M}%
}{1-\frac{2\epsilon}{M}}\right)  \right] \nonumber\\
&  =\left(  2n+1\right)  4\pi M\omega
\end{align}
where we have written $(r-M)=-\epsilon e^{i\theta}$ and $n$ is an integer. The
first part of the integral is a real contribution of $O(\frac{1}{\epsilon}%
)$\ that diverges as $\epsilon\rightarrow0$. \ It does not contribute to the
imaginary part of the action. \ \ The imaginary part of the action leads to a
non-zero finite temperature \
\begin{equation}
T_{H}=\frac{1}{8\pi M(2n+1)} \label{unphys}%
\end{equation}
for any integer $n$. \ The extremal temperature is quantized in units of the
temperature of a Schwarzschild black hole!

Note that this result depends crucially on the inclusion of the third order
term, whose evaluation depends upon assumptions of the choice of Riemannian
sheet. \ Had we expanded the integral for small $\epsilon$ , we would have
obtained a value for the temperature given by $n=-1$ in eq. (\ref{unphys}), ie
a negative temperature for the extremal black hole.

Obtaining many (finite-valued) results for the temperature is reminiscent of
the proposal that an extremal black hole can be in thermal equilibrium at any
finite temperature \cite{HHR}. \ However we can see that these strange results
arise due to an inappropriate use of the WKB approximation in the null
geodesic method. \ Although writing $(r-M)=-\epsilon e^{i\theta}$ is
consistent with the the assumptions $r_{in}=r_{0}(M)-\epsilon$\ and
$r_{out}=r_{0}(M-\omega)+\epsilon$\ (where $r_{0}(M)$\ denotes the location of
the event horizon of the original background space-time) for a non-extremal
black hole, in fact the quantity $r_{out}$ does not exist, since the extremal
black hole cannot retain an event horizon upon emitting any neutral quantum of
energy -- its only option for future evolution would appear to be that of
evolving into a naked singularity, which cosmic censorship forbids. \ 

These results seem to imply that for black holes near extremality one must
consider the full self-gravitating results, where the emitted particle drives
the hole toward extremality. \ For an already extremal spacetime both methods
yield a diverging real component in the action. This could be taken to imply
that no particle can be emitted (since the alternative is creation of a naked singularity).

Based on the results of this calculation it would be interesting to consider
the emission of a specific charged particle that would cause the black hole to
go from one extremal black hole to another extremal black hole. \ In that case
there would be well defined horizons before and after emission. \textbf{\ }

\section{Conclusions}

We have examined and compared the two different approaches to the tunnelling
method for finding the black hole temperatures. \ Our results indicate that
the method is particularly robust for non-extremal black holes, yielding
results commensurate with other methods for Rindler space, rotating black
holes, and Taub-Nut black holes. In this latter instance we have provided
independent verification of the temperatures obtained for Taub-NUT spaces
without CTCs via analytic continuation methods. Indeed it is not too difficult
to show that the temperatures even match when CTCs are present, though in this
case an a-priori justification for the method is unclear.

We also investigated extremal black holes, for which the tunnelling method is
somewhat more problematic due to its dynamic nature. We found that the
temperature is proportional to the energy of the emitted particles for black
holes close to extremality. \ We also found that both methods yield a
divergent real part to the action for extremal black holes, which is
suggestive of a full suppression of particle emission. However the null
geodesic method has a nonzero finite imaginary parts, whose values yields a
countably infinite number of possible finite temperatures for an extremal
Reissner-Nordstrom black hole. \ This rather strange result arises because of
a breakdown of the WKB method in the null geodesic approximation. This
suggests limitations on the method, whose study would make an interesting
subject for future work. An interesting test case would be that of emission of
charged particles from an extremal black hole.

\section{Appendix}

\subsection{CTC's and Taub-NUT space}

\ The presence of closed timelike curves in Taub-NUT space can be seen by
considering the curve generated by the Killing vector $\partial_{\varphi}%
$\ and by examining $g_{\varphi\varphi}$
\[
g_{\varphi\varphi}=4f_{k}^{2}(\frac{\theta}{2})\left(  r^{2}+n^{2}-f_{k}%
^{2}(\frac{\theta}{2})(4n^{2}F+k(r^{2}+n^{2}))\right)
\]
So for $k=1,0$, and $k=-1$ with $4n^{2}/\ell^{2}>1$ the quantity
$g_{\varphi\varphi}<0$, yielding a timelike $\partial_{\varphi}$; the curve
$r=r_{0}$, $t=t_{0}$, and $\theta=\theta_{0}$ becomes a CTC.

\ However there is a range of hyperbolic Taub-NUT solutions that occur when
$4n^{2}/\ell^{2}\leq1$ that don't contain CTC's. \ Now it is possible for
$g_{\varphi\varphi}$ to be negative when $4n^{2}/\ell^{2}<1$but this occurs
for small values of $r_{0}$ and happens inside the horizon. Explicitly when
$k=-1$ then $g_{\varphi\varphi}$ is given by%
\[
g_{\varphi\varphi}=4\sinh^{2}(\frac{\theta}{2})(r^{2}+n^{2})\left(  \cosh
^{2}(\frac{\theta}{2})-\frac{4n^{2}F}{r^{2}+n^{2}}\sinh^{2}(\frac{\theta}%
{2})\right)
\]
So $g_{\varphi\varphi}\geq0$ will always be true as long as $\frac{4n^{2}%
F}{r^{2}+n^{2}}\leq1.$ Figures 1-3 are plots of $1-\frac{4n^{2}F}{r^{2}+n^{2}%
}$\ for a range of mass and NUT-charge . \ On the plots the x-axis is $r/n$.
The $k=-1$\ case corresponds to hyperbolic solutions whose event horizon has
radius $r_{b}>n$. Since $g_{\varphi\varphi}$\ only becomes negative when
$r<n$\ (within $4n^{2}/\ell^{2}\leq1$) any CTCs are contained within the
horizon (provided the mass is positive). \ So no CTC's are present outside of
the horizon for the hyperbolic case when $4n^{2}/\ell^{2}\leq1.$%
%TCIMACRO{\FRAME{dtbpFU}{3.314in}{3.1687in}{0pt}{\Qcb{Figure 1: Plots when
%4n$^{2}$/L$^{2}<1$ for a range of masses}}{\Qlb{figure1}}{plot1.eps}%
%{\special{ language "Scientific Word";  type "GRAPHIC";  display "ICON";
%valid_file "F";  width 3.314in;  height 3.1687in;  depth 0pt;
%original-width 5.5564in;  original-height 5.5564in;  cropleft "0";
%filename 'plot1.eps';file-properties "XNPEU";}}}%
%BeginExpansion
\begin{center}
\includegraphics[
height=3.1687in,
width=3.314in
]%
{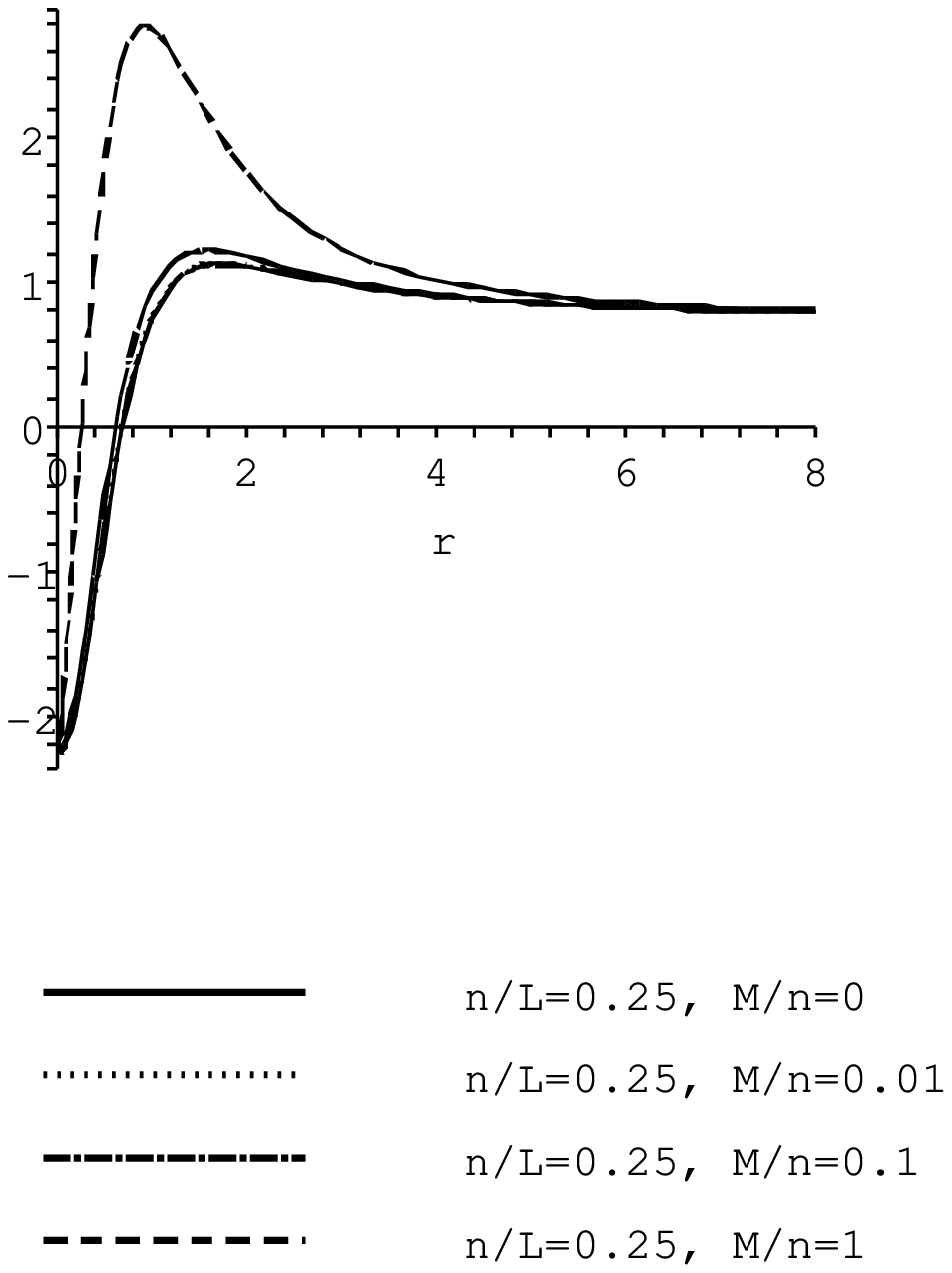}%
\\
Figure 1: Plots when 4n$^{2}$/L$^{2}<1$ for a range of masses
\label{figure1}%
\end{center}
%EndExpansion%
%TCIMACRO{\FRAME{dtbpFU}{2.9456in}{3.1678in}{0pt}{\Qcb{Figure 2: Plots for when
%4n$^{2}/$L$^{2}=1$ for range of masses.}}{\Qlb{Figure2}}{plot2.eps}%
%{\special{ language "Scientific Word";  type "GRAPHIC";  display "ICON";
%valid_file "F";  width 2.9456in;  height 3.1678in;  depth 0pt;
%original-width 6.0502in;  original-height 6.7343in;  cropleft "0";
%filename 'plot2.eps';file-properties "XNPEU";}}}%
%BeginExpansion
\begin{center}
\includegraphics[
height=3.1678in,
width=2.9456in
]%
{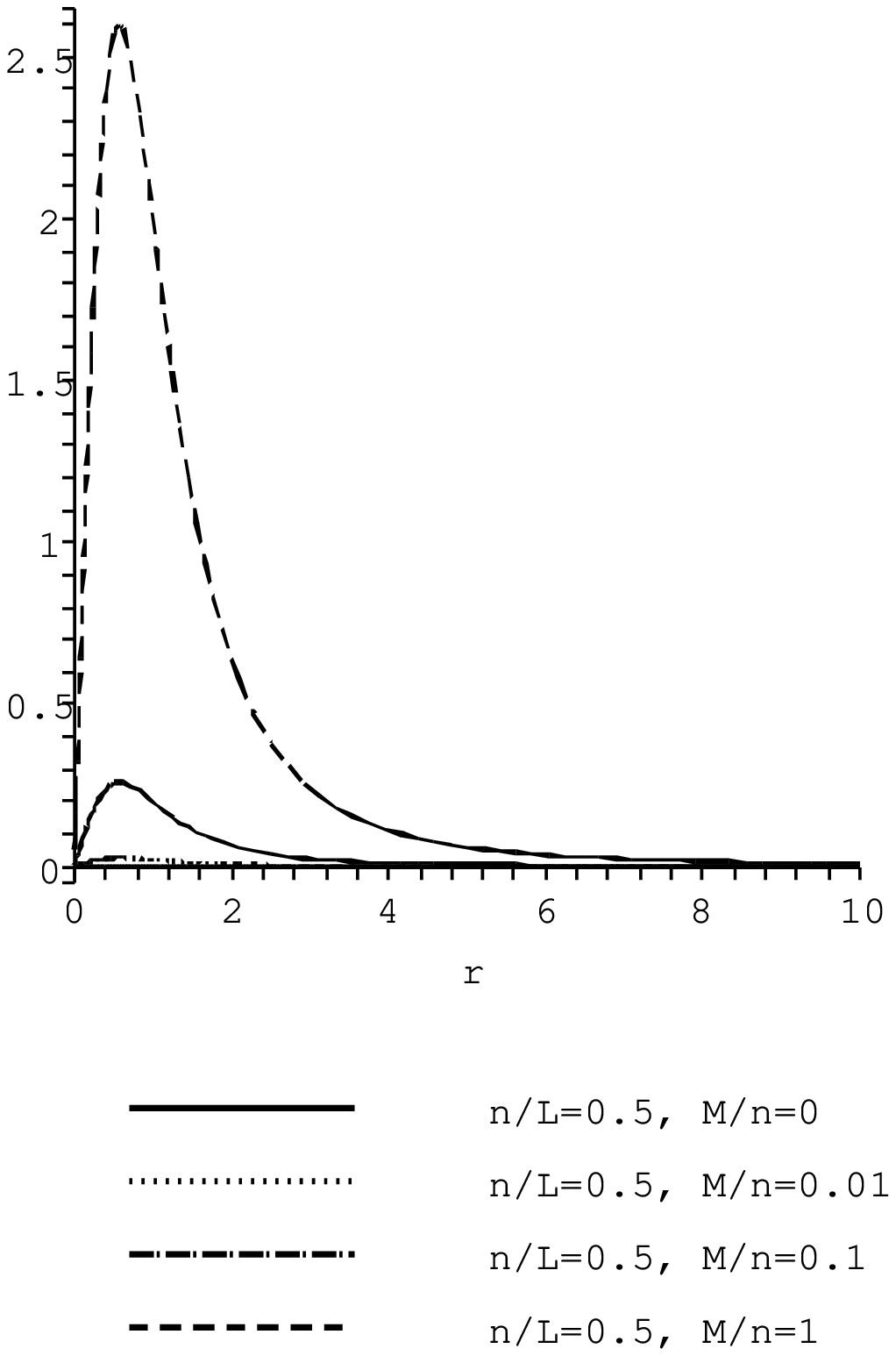}%
\\
Figure 2: Plots for when 4n$^{2}/$L$^{2}=1$ for range of masses.
\label{Figure2}%
\end{center}
%EndExpansion%
%TCIMACRO{\FRAME{dtbpFU}{4.2627in}{2.4137in}{0pt}{\Qcb{Figure 3: Plots for
%fixed mass and a range of n$^{2}/$L$^{2}$}}{\Qlb{Figure3}}{plot3.eps}%
%{\special{ language "Scientific Word";  type "GRAPHIC";  display "ICON";
%valid_file "F";  width 4.2627in;  height 2.4137in;  depth 0pt;
%original-width 7.2774in;  original-height 5.4232in;  cropleft "0";
%filename 'plot3.eps';file-properties "XNPEU";}}}%
%BeginExpansion
\begin{center}
\includegraphics[
height=2.4137in,
width=4.2627in
]%
{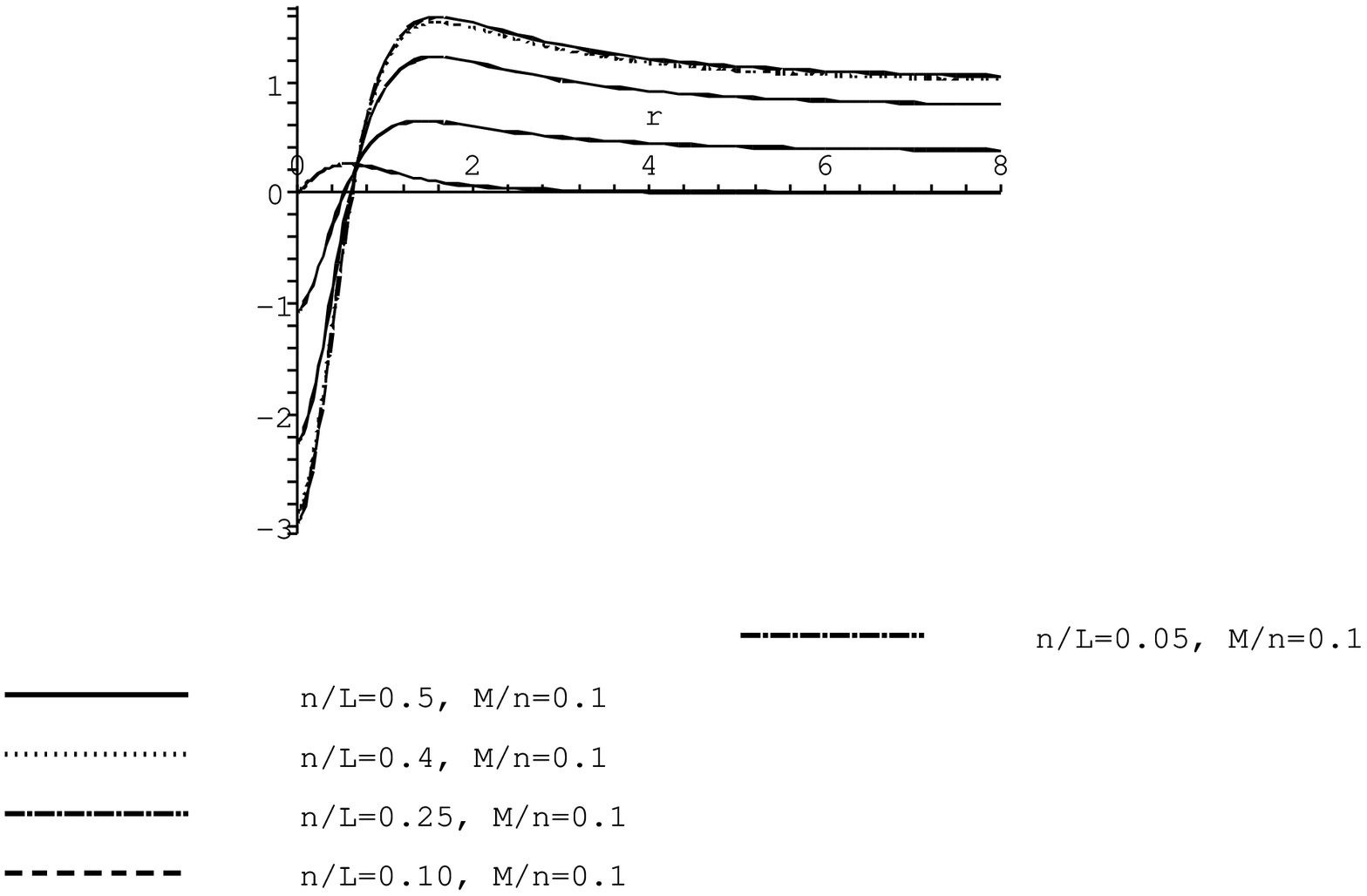}%
\\
Figure 3: Plots for fixed mass and a range of n$^{2}/$L$^{2}$%
\label{Figure3}%
\end{center}
%EndExpansion

\bigskip

{\Large Acknowledgements}

This work was supported by the Natural Sciences and Engineering Research
Council of Canada. We are grateful to Maulik Parikh and Per Kraus for
interesting discussions and correspondence concerning this work.

\end{document}